# CHARACTERIZATION OF ALL-CHROMIUM TUNNEL JUNCTIONS AND SINGLE ELECTRON TUNNELING DEVICES FABRICATED BY DIRECT-WRITING MULTILAYER TECHNIQUE

H. Scherer, Th. Weimann, P. Hinze, B.W. Samwer, A.B. Zorin, and J. Niemeyer

*Physikalisch-Technische Bundesanstalt, Bundesallee 100, D-38116 Braunschweig, Germany*

**Abstract**

We report about the fabrication and analysis of the properties of $Cr/CrO_x/Cr$ tunnel junctions and SET transistors, prepared by different variants of direct-writing multilayer technique. In all cases, the $CrO_x$ tunnel barriers were formed in air under ambient conditions. From the experiments on single junctions, values for the effective barrier height and thickness were derived. For the $Cr/CrO_x/Cr$ SET transistors we achieved minimal junction areas of 17 x 60 $nm^2$ using a scanning transmission electron microscope for the e-beam exposure on $Si_3N_4$ membrane substrate. We discuss the electrical performance of the transistor samples as well as their noise behavior.

## 1. Introduction

The growing interest in the physics and application of devices based on single electron tunneling (SET) effects in the present decade strongly calls for the maximal possible reduction of the structure dimensions in order to increase the Coulomb charging energy $E_c = e^2/2C$, where $e$ is the electronic charge and $C$ the characteristic capacitance parameter of the structure, and thus, by virtue of the necessary condition of $E_c \gg k_B T$, to increase the operating temperature $T$ or to improve the general performance of SET devices.

The known methods suitable for the preparation of SET systems are manifold. There are, for instance, different techniques utilizing a scanning probe microscope in various ways (see e.g. [1-3]). Also, the methods utilizing small clusters of granular metallic films arouses great interest.[4] But the potential of the most prominent family of technologies, namely the methods based on e-beam lithography (EBL), at present still seems to be unsurpassed. Despite the fact that the methods mentioned first have successfully been applied for the preparation of simple devices like SET transistors, up to now, the mature and fast EBL-based technologies seem to be best qualified for the fabrication of more complex SET circuits like electron turnstiles, pumps or potential logic devices and memories (see e.g. [5-8]). Also, EBL seems to be the most promising way to integrate SET devices into other circuits. Therefore, the development and improvement of EBL techniques is very important for the future development of SET applications.

Diverse fabrication techniques based on EBL have been established,[9,10] but the EBL variants based on the shadow evaporation method still represent the world-wide standard for the fabrication of SET circuits. This method allows the *in situ* fabrication of small, high-quality tunnel junctions using only one resist mask.[11,12] The main disadvantage is the unavoidable occurrence of stray shadows, complicating the layout of the circuit. This is the reason for the development of direct-writing multilayer techniques. Here, the different layers of the metallic SET circuit are deposited in separate process steps using individual resist masks. Thus, the flexibility of the circuit layout is enhanced owing to the lack of parasitic features and, in addition, the combination of different materials for each layer is naturally provided.

Recently, the fabrication and operation of a metallic multilayer SET transistor has been demonstrated, the charge island having the shape of a suspended bridge.[13] In this work, we present a simpler, straightforward fabrication technique, allowing the multilayer preparation of all-chromium ($Cr/CrO_x/Cr$) tunnel junctions and SET transistors.

The paper is organized as follows: In chapter 2, we briefly discuss the suitability of the all-chromium system for the multilayer technique. In part 3, we describe the preparation of $Cr/CrO_x/Cr$ tunnel junctions and present an analysis of the barrier properties, including the height and thickness. In chapter 4, we demonstrate that the multilayer technique can be used for direct-writing fabrication of $Cr/CrO_x/Cr$ SET transistors. We briefly sketch the different preparation variants, i.e. the preparation by exclusively "conventional" EBL on bulk substrates, and a new method based on the combination of "conventional" EBL and e-beam exposure by a scanning transmission electron microscope (STEM) on an $Si_3N_4$ membrane substrate. We investigate and discuss the electric performance and the noise spectra of these SET transistors and finally summarize the results.

## 2. Technological and material aspects

For a number of reasons which will be briefly discussed in the following, the task of developing multilayer methods for SET devices is directly linked with material questions. Unfortunately, the standard material for SET devices, aluminum, is not favorable for multilayer methods for several reasons: Indeed, it should be possible for the material to be deposited on the substrate in very thin (say some 10 nm), but continuous, conducting layers. The requirement for small film thickness results from the fact that for SET devices the lateral dimensions aimed at usually are in the region below 100 nm.[14] Furthermore, it is necessary for the deposited metal films to remain conductive though they are exposed to air, since the vacuum condition usually is interrupted between two lithographic steps of a multilayer process. As to aluminum, it is known that very thin (< 20 nm) aluminum films tend to become insulating when exposed to air.[15] Also, it is reported that the spontaneously formed $AlO_x$ surface layer is about 5 nm thick,[16] which would lead to almost vanishing tunnel conductance for small $Al/AlO_x$/metal tunnel junctions. To circumvent these problems, complicated passivation techniques have to be applied if multilayer structures are fabricated using aluminum electrodes.[13]

Instead, looking for alternative materials, the metals should preferably be chemically inert or, even better, self-passivating by forming a very thin insulating layer on its surface (for instance, a native oxide of the metal), which could also be used as a tunnel barrier. Chromium or titanium for example are suitable candidates, since they have proved their worth in thin-film technology and meet the above-mentioned requirements. Moreover, chromium and titanium films are known for their excellent adhesion to substrate materials; they are fine-grained and, thus, allow very smooth surfaces and finest structures to be defined.[15,17] The high melting points and the hardness of the two materials indicate high (thermal) stability of the thin-film structures. Their surface oxide layers, spontaneously formed at room temperature, remain thin enough to serve as natural tunnel barriers. For chromium, the stoichiometry of the air-formed surface oxide layer is reported to be close to $Cr_2O_3$,[18,19] but generally at least three different oxides may be formed under normal conditions ($Cr_2O_3$, $CrO_2$, $CrO_3$).[20] For the saturated thickness of chromium oxide layers formed under ambient conditions, several authors report values in the range between 0.9 nm and 2.0 nm.[18-23] It is known that titanium also forms a thin coating layer,[24,25] but the composition seems to depend on the conditions of preparation and is controversially discussed in the literature. Under ambient conditions, $TiO_2$ seems to be the main oxide phase[26,27] with a saturated thickness of ≈ 1 nm.[15,28]

Both material systems have recently proved to be suitable for the preparation of SET devices using shadow evaporation techniques.[15,21,22] Nevertheless, when comparing the dielectric constants of $Cr_2O_3$ ($\varepsilon \approx 10 - 12$) and $TiO_2$ ($\varepsilon > 90$, room temperature values) we believe that chromium is superior to titanium.[29,30] The latter value even increases at lower temperatures, which seems to be a fatal drawback for $TiO_2$ in SET applications because of the undesired increase in the tunnel junction capacitance. Also, the attractive combination of SET devices with small-sized, but high-ohmic chromium thin-film resistor components (see e.g. [31]) might be simplified when an all-chromium system is used.

## 3. Characterization of $Cr/CrO_x/Cr$ tunnel junctions

Generally, the literature about $Cr/CrO_x$/metal junctions is sparse, but recently several approaches to the fabrication of chromium-based SET transistors have been reported.[21,22,32] In two of those works attempts are made to characterize the tunnel properties of $Cr/CrO_x/Cr$ junctions, but the $CrO_x$ barrier formation was achieved by *in situ* oxidation[21] or direct $CrO_x$ evaporation,[22] which is quite different from our process in which the barrier was formed in air under ambient conditions. In both cases, the junctions were characterized by fitting of the Simmons model for the tunneling current to the measured current-voltage characteristics (IVCs). This simple model, valid for plate-capacitor-like electrodes of similar type, parameterizes the tunneling current by the (effective) height $\phi$ and width $d$ of the (rectangular) tunnel barrier.[33] Both authors find $\phi$ = 170 meV - 175 meV and $d$ = 1.6 nm - 2.0 nm, but their results seem debatable since in both cases the quality of the fits was poor. As Pashkin *et al.* pointed out,[22] the agreement between experiment and theory at low temperature especially suffered from a strong non-linearity of the measured IVCs at low bias. In other words, he found a drastic increase in the differential resistance in the low voltage regime, measured on a single $Cr/CrO_x/Cr$ junction, that could not be explained within the Simmons model.

For closer re-investigation of the general tunnel behavior of $Cr/CrO_x/Cr$ junctions and for characterization of our tunnel elements, made by the direct-writing process, we intended to perform an analysis on the basis of single junctions. Bottom and counter electrodes were prepared in the simple way by crossing two chromium lines, and the geometrical areas $A$ of the junctions were about 0.5 x 0.5 µm² and 1 x 1 µm². These relatively high values were chosen since the characterization was carried out on the basis of current-voltage characteristics, and we wanted to ensure that the "normal" tunneling behavior was not masked by SET effects.[34]

*Sample preparation*

For all EBL steps, we used a Leica EBPG-4 HR electron beam writing system (50 keV) and a two-layer e-beam resist of PMMA 200k 3% / PMMA 950k 1% (bottom / top layer), providing the necessary undercut for good lift-off properties, and a solution of isopropyl alcohol (IPA): methylisobutylic ketone (MIBK) (3:1) + 1% methylethylketone (MEK) as developer system.[35] The substrate was a silicon wafer with a thermally oxidized $SiO_2$ surface (600 nm thick). First, the markers (10 µm wide squares of Cr/Au) for the adjustment of the different resist masks were prepared by a lift-off process. Next, the resist mask for the bottom electrodes and the contact pads was prepared. These structures were made by a lift-off step of a 15 nm thin chromium film, deposited by thermal evaporation at a rate of 0.2 nm/s and a background pressure of about $10^{-4}$ Pa. Before opening the shutter for deposition, some chromium was evaporated to degas the material and to take advantage of its excellent getter properties. After lift-off, the wafer was stored in the clean-room under ambient conditions ($T = 20$ °C and about 50 % relative humidity) for about 24 hours to allow the insulating surface layer to be formed. Then, the next resist mask was prepared and the 35 nm thick chromium top electrodes (perpendicularly crossing the bottom electrodes) and contact pads again were defined by lift-off. A minimum thickness of about 30 nm for this second layer was necessary to avoid cracks at the edges of the bottom structures. Finally, the contact pads were covered by a gold film (20 nm thick) to ensure reliable contact properties.[36] Both ends of the electrode strips were contacted to carry out four-terminal measurements across the junction. The sheet resistances of the chromium films scaled with their thickness and were about 94 Ω per square (37 Ω per square) for the 15 nm (35 nm) thick films (at $T = 4.2$ K). Preliminary investigations of the tunnel junctions showed a moderate process yield (about 75 % of the junctions were not electrically open or shorted), but a considerable spread of their resistance values over the whole wafer. Also, on many chips we found that the junction resistances did not scale properly with the geometrical tunnel areas. However, for the following experiments, sample chips were chosen showing reasonable resistance scaling of their differently sized junctions.

*Experiment*

The custom-made electronic setup for the characterization of the single junctions was essentially the same as for the experiments on the SET transistor samples. All electric measurements were carried out symmetrically with respect to ground. The symmetric bias voltage was connected to high-ohmic loading resistors, biasing the sample with a constant current. The resistors and a low noise preamplifier to measure the voltage across the tunnel junction were installed in a closed screening box mounted directly on top of the sample holder. Since the further analysis described below required the application of rather high voltages across the junctions (up to about 1 V), we recorded these IVCs in the liquid helium bath to achieve the most effective cooling of the samples. To exclude any irreversible changes of the junction behavior due to the high electric field across the barrier, we checked the reproducibility of the IVCs during the measurements.

*Analysis*

Figure 1 shows IVCs recorded at room temperature, in a liquid nitrogen bath ($T = 77$ K) and in a liquid He bath ($T = 4.2$ K). The arrows point to regions of an IVC where spurious events occurred that looked like multi-level switching of the IVC between different branches. Similar observations were made on most of the samples investigated, also at helium temperatures, where the switching events mainly appeared at higher bias voltages. We attribute this behavior to impurities in the tunnel junction located inside the tunnel barrier or, by contamination of the $CrO_x$ surface after oxidation, on the interface between tunnel barrier and metal electrode. Probably these impurities (e.g. carrier traps) are spontaneously activated by the high electric field in the barrier, switching additional current paths on and off. As a general trend for all samples measured, we found that the junction resistances rose with decreasing temperature, probably due to the suppression of thermally activated transport via the tunnel barrier.

Plotting the differential resistance (inset of Figure 1), we found an unusually sharp peak in the low bias region. This strong non-linearity of the IVCs around zero bias behavior is anomalous in the sense that, according to the basic theories for tunneling,[33,37] the tunneling current expression for "normally" behaving metal-insulator-metal junctions can be expanded like $I(U) = aU + bU^3 + ... \; (a, b \neq 0)$, and should, thus, be approximately linear around zero bias. The low bias anomaly was observed on all samples investigated and seems to be typical of chromium-based junctions, as also the recent work by Pashkin *et al.* suggests.[22] Although a definite explanation for this phenomenon does not exist at present, we point out that in qualitative terms it is similar to the observations of Rowell and Shen on Cr/CrO$_x$/Ag and Cr/CrO$_x$/Pb junctions.[38] These authors attribute the appearing zero bias tunneling anomalies to the magnetic nature of chromium and the barrier oxide. Therefore, at least for the low bias range, it seems obvious that the measured IVCs cannot adequately be explained in the framework of the simple Simmons model. Instead, the theoretical description of the complex tunneling behavior of chromium-based junctions will probably require more complex models which are not, however, available at present. Nevertheless, in the high bias range, the anomaly effect should be less pronounced and the curvature of the IVC should rather be dominated by the barrier suppression effect. To test this hypothesis, we tried to model the curvature of the measured IVCs in the large bias region using the Simmons formula[33]

$$j = \frac{e}{2\pi h d^2} \cdot$$

$$\left\{ \left( \phi - \frac{eU}{2} \right) \exp\left[ -\frac{4\pi d}{h} \sqrt{2m} \sqrt{\left( \phi - \frac{eU}{2} \right)} \right] \right.$$
$$\left. - \left( \phi + \frac{eU}{2} \right) \exp\left[ -\frac{4\pi d}{h} \sqrt{2m} \sqrt{\left( \phi + \frac{eU}{2} \right)} \right] \right\}. \quad (1)$$

Here, $j$ is the tunnel current density for $|U| < \phi/e$, $h$ Planck's constant, $m$ the free electron mass, and $d$ and $\phi$ are the fitting parameters for the effective barrier thickness and the effective height. Since (1) is valid for $T = 0$, we took into account the modifications for the case of finite temperature,[39] but the effect of this correction appeared to be very small at 4.2 K. To evaluate only the shape of the IVCs, we normalized the curves before comparison. After $d$ and $\phi$ had been obtained, the parameter $A^*$ for the effective tunnel area was adjusted to fit the measured absolute current value $I = A^* j$.

The results of the analysis for two differently sized junctions are summarized in Table 1, and a corresponding plot of the measured and the fitted curve for the sample with $A = 0.53 \times 0.53$ µm² is given in Figure 2. Before turning to the discussion of the results for the fitting parameters, we point out that we obtained very good agreement between theory and experiment for the high voltage range, i.e. for $|U| > 400$ mV, as the IVCs in Figure 2 and their differential resistances (inset) show. In the lower bias range, the experimental curve is shifted to higher voltages with respect to the model. The quality and the features of the corresponding plot for the second junction (not shown) were similar. As the table shows, the values derived for the effective thickness of the tunnel barrier are about $d = 1.35$ nm – 1.40 nm.

This value is reasonable for a CrO$_x$ barrier grown at ambient temperature and fits well in the range of values reported by others.[18-23,40] The ratios between the effective and the geometrical junction areas $A^*/A$ were about 10 % – 20 %. In the literature, other works sometimes reported similar observations, for instance on Al/AlO$_x$/Al, Cr/CrO$_x$/Pb and V/VO$_x$/Pb tunnel junctions it was found that only a small fraction of the area carried an appreciable part of the current.[40] As the most probable explanation for this behavior, we consider a partial contamination of the bottom electrode surface with residues of e-beam resist or other chemicals caused by our preparation process. Also, a non-uniform thickness and/or non-uniform oxidation of the CrO$_x$ layer might lead to current concentration on small parts of the junction area.[41]

The most interesting point is the value of $\phi \approx (740 \pm 50)$ meV for the effective barrier height, which is considerably larger than $\phi \approx 170$ meV as derived by Kuzmin et al.[21] and Pashkin et al.[22] Nevertheless, our result is confirmed by two important points: First, the quality of our fit in the higher bias regime is more convincing in comparison with the earlier analysis presented in [22]. This is caused by the different criteria for the fitting procedure: In our case, we aimed at good agreement in the higher bias regime, whereas in the earlier works the whole bias range was taken into account, i.e. also the low-bias region where the fitting of (1) had to fail.

A second hint for a higher barrier value is given by the further analysis of the IVCs at high bias voltages. It is well kown that for $U >> \phi/e$ (i.e. the Fowler-Nordheim or field-emission tunneling regime) the tunneling current follows the relation[42]

$$I \propto U^2 \exp\left( -\tfrac{2}{3} \alpha d \phi^{3/2} / eU \right) \quad (2)$$

with $\alpha = 2(2m)^{1/2} / \hbar$.

Thus, in the high bias region, the plot $\ln(I/U^2)$ against $1/U$ (Fowler-Nordheim plot) should asymptotically approach a straight line with negative slope equal to $-d\phi^{3/2}$. Such a plot is shown in Figure 3 for a Cr/CrO$_x$/Cr junction with $A = 1 \times 1$ µm². It is obvious that here, even at a maximum voltage of more than 1 V, the Fowler-Nordheim regime was not reached, the curves rather show only weak and non-linear bias dependence. Unfortunately, a further extension of the bias range was not possible since all junctions investigated became unstable, irreproducibly changed their IVC and/or were destroyed above $\approx 1$ V. Although the slope in the field-emission regime could not be determined experimentally, Figure 3 allows the minimum value of the barrier height to be estimated: The slope of the curves at the highest bias voltage was about $-3.2$ V$^{-1}$ (indicated by the short dashed line in Figure 3), actually underestimating the expected slope of the asymptote in the Fowler-Nordheim regime. Hence, $d\phi^{3/2} \geq 3.2$ V$^{-1}$, and on the realistic assumption of a maximum value $d_{max} = 2$ nm for the barrier thickness (cf. chapter 2), we may deduce a minimum value $\phi_{min} = 380$ meV.

Thus, our analysis gives evidence that the barrier height is at least about twice as large as the value stated in [21] and [22]. We conclude that the tunneling behavior of Cr/CrO$_x$/Cr junctions is dominated by a zero-bias anomaly in the low bias regime, not understood at present, and that previous approaches to determining the barrier parameters by fitting (1) suffered from the inadequate choice of the parameter range. In an extended bias range, formula (1) seems to be an adequate approximation to the tunneling current, being dominated by the barrier suppression at larger voltages. The fitted value of $\phi \approx 740$ meV is about twice to three times smaller than the value for traditional Al/AlO$_x$/Al junctions.[40] Due to the inherent non-linearity of the "seed" IVC of Cr/CrO$_x$/Cr junctions around zero bias, all SET effects studied in the following appeared on a non-linear background. Nevertheless, as we will see, this peculiarity of chromium-based SET devices did not spoil the functionality of our SET transistors.

## 4. Experiments on all-chromium SET transistors

*a) SET transistor prepared by "conventional" EBL*

The transistor type shown in Figure 4 was prepared in the same run and on the same wafer as the Cr/CrO$_x$/Cr single junctions described in the previous section. In the first lift-off step, the source, drain and gate electrodes were structured from the bottom chromium layer, followed by the 24 h oxidation under ambient conditions. A crossing line forming the island was made in a subsequent lift-off step. According to the linewidths of the island and electrode strips, the area of the Cr/CrO$_x$/Cr tunnel junctions was about 100 x 100 nm$^2$, and the total island length was about 6.5 µm.

Experiments were carried out in a top-loading dilution refrigerator with a base temperature of 25 mK. All signal lines were equipped with commercial Thermocoax® cables about 1 m in length, serving as microwave frequency filters.[43] Figure 5 (a) shows IVCs of the transistor in different bias regions and at different gate voltages $V_g$. Due to the low bias tunneling anomaly which is not SET-related and the barrier suppression effect in the high voltage range, the IVC of the transistor is non-linear in the whole bias range (see inset). Gate modulation was observable up to a temperature of $\approx 1$ K.

Figure 5 (b) shows gate modulation curves at different values of bias currents, giving a gate capacitance $C_g \approx 53$ aF. A capacitance ratio of $C_1/C_2 \approx 0.86$ for the tunnel junctions (including stray capacitance contributions) was determined from the slopes of the approximately triangular modulation curves. We also measured the current modulation at constant voltage bias (not shown) and determined the ratio $R_2/R_1 \approx 0.84$ for the tunnel resistances in the standard way (see e.g. [44]). Hence, $R_1C_1 \approx R_2C_2$, implying that both tunnel junctions were of the same thickness in this particular sample.

For aluminum-based SET transistors, the determination of the total island capacitance $C_\Sigma$ is usually carried out by analysis of the offset plot $U_{\text{Offset}} = U - I(dI/dU)^{-1}$ vs. $U$, i.e. by evaluation of the linear asymptotes of the IVC.[45] In the case of chromium-based transistors, this method is not feasible because of the overall non-linearity of the seed IVC. Instead, we estimated $C_\Sigma$ from modulation curves recorded at low temperatures and small currents $I = \pm 5$ pA. These curves are shown in the inset of Figure 5 (b) and allowed the approximate shape of the stability diagram for the transistor (gray rhomb) to be restored. Using $C_\Sigma = e/U_{\text{max}}$, we found $C_\Sigma \approx (820 \pm 40)$ aF from the corresponding maximum modulation amplitude $U_{\text{max}} \approx 195$ µV. With these values, using $C_\Sigma = 2 C_T + C_g + C_0$, we obtained $C_T \approx 300$ aF for the junction capacitance. For the self-capacitance $C_0$ of the rather long island, a value of $C_0 \approx 170$ aF was estimated by means of the relation[45,46]

$$C_0 = 2\pi\varepsilon_0\varepsilon_{eff}\frac{l}{\ln(8l/w)} \qquad (3)$$

(island strip length $l = 6.5$ µm and width $w = 100$ nm). $\varepsilon_{eff} = 3$ was plugged in for the effective dielectric constant of the surroundings, an arithmetic average of values of the substrate material underneath the island (SiO$_2$, $\varepsilon \approx 5$) and of liquid helium ($\varepsilon \approx 1$). Assuming that $C_T$ can be approximated by the simple plate capacitor formula (with $A = 0.01$ µm² and $d = 1.4$ nm according to the result derived from the single junctions), we calculated $\varepsilon_r \approx 4.5$ for the dielectric constant of the barrier material. Although this derived $\varepsilon_r$-value is significantly smaller than expected for pure Cr$_2$O$_3$ ($\varepsilon_r \approx 10 - 12$),[29,30] similar observations of effectively reduced dielectric constants in tunnel barriers are sometimes reported by others (for instance for AlO$_x$ barriers, see [47]). On the other hand, this might indicate that the tunnel barrier is contaminated or composed of different materials. Unfortunately, no data for the dielectric constants of other chromium oxide phases were found in the literature for comparison, however, we found that the dielectric constant of PMMA is $\varepsilon_r \approx 3$. Therefore, a possible contamination with a thin residual film of e-beam resist of parts of the bottom electrode cannot be ruled out. Clarification of the junction quality and homogeneity will require further investigations of transistor samples with differently sized junctions and islands.

Figure 6 shows noise spectra for 1 Hz $< f <$ 100 Hz, measured in the current bias mode. First, as shown in Figure 6 (a), we investigated the voltage noise behavior in different working points of the modulation curve, i.e. in the maximum (A) and minimum (C) of $U(V_g)$, corresponding to vanishing charge sensitivity, and in point (B) of maximum slope and charge sensitivity. In the first case, we found that the voltage noise was close to the noise floor of our electronics setup of about 30 nV/√Hz at 10 Hz and exhibited $1/\sqrt{f}$ frequency dependence.[48] Since in these working points the transistor is merely sensitive to fluctuations of the tunnel junction resistance, we conclude that such fluctuations are weak and the measured spectrum is dominated by the noise of our preamplifier (see also [49]). In working point (B) (maximum charge sensitivity) we observed an increased voltage noise, indicating that the origin of the noise are the well-known fluctuations of background charges in the vicinity of the transistor island. For Figure 6 (b) we converted the voltage noise spectra, measured at different bias currents, into the equivalent charge noise spectra, using the relation $Q_{Noise} = C_g U_{Noise}/(dU/dV_g)$. Usually, background charge noise is generated by many independently and randomly switching two-level-fluctuators (TLFs). The spectrum of a single TLF is of Debye-Lorentzian shape, whilst the superposition of the contributions of many TLFs results in a total spectrum with the usually observed $1/f$ dependence (see e.g. [50] and references therein). From Figure 6 (b) we see that for small currents the spectra approximately follow a $1/f$ dependence, but change to a $1/\sqrt{f}$-like behavior at higher currents. An explanation for this might be the current-dependent activation of TLFs: For small currents only a few TLFs are active, so that the $1/f$-like regime of their individual Debye-Lorentzian shaped spectra dominates the total spectrum. For higher currents, more TLFs become activated, and finally the $1/\sqrt{f}$-like spectrum of many individual switching traps develops. The physical mechanisms for this activation might be either the electric potential fluctuations of the transistor island due to the shot noise–like tunneling events and/or the local overheating of the dielectric surrounding of the island.[51-53]

As can be seen from Figure 6 (b), the sample showed a moderate noise figure of only about 2.5 x 10$^{-4}$ e/√Hz at $I$ = 5 pA and $f$ = 10 Hz, monotonically increasing with the bias current up to ≈ 1.5 x 10$^{-3}$ e/√Hz at 150 pA. Such a behavior is quite similar to that of traditional Al/AlO$_x$/Al transistors prepared on similar substrates,[52] and, therefore, likely reflects noise properties of the substrate material in the vicinity of the transistor island rather than intrinsic properties of individual transistor samples.

*b) SET transistor prepared by STEM lithography*

The SET transistor shown in Figure 7 was prepared on a membrane of Si$_3$N$_4$ (20 nm thick). This special substrate had to be used since we utilized a scanning transmission electron microscope (STEM, Philips CM 200 FEG with ELPHY 1 writing system) to define the central transistor island. First, we structured the source, drain and gate electrodes together with position markers and contact pads by a "conventional" Leica EBPG-4 HR writer. After deposition, lift-off and oxidation of the bottom chromium layer, a crossing line of about 10 µm in total length was written using the STEM. This line (about 17 nm wide) was made of a double metal layer of chromium (5 nm) with gold/palladium (15 nm) on top, and served as the transistor island. The purpose of the Au/Pd layer on top of the island was to fill interruptions possibly appearing in the line. With the exception of this point, all lithographic steps were performed in a way similar to that described in the previous section. Determined by the overlapping areas of the island and the electrode lines, very small Cr/CrO$_x$/Cr tunnel junctions with areas of ≈ 17 x 60 nm$^2$ were formed (see inset). A more detailed description of the fabrication procedure can be found elsewhere.[54]

Experiments were carried out in a dilution refrigerator with a base temperature of 8 mK. In contrast to the former experiment, in this fridge, the sample was mounted in a closed screening box. The filtering of the electrical lines and the electronics setup were similar to those described above.

Figure 8 (a) shows IVCs with the clear Coulomb blockade of the transistor. Gate modulation with a period corresponding to $C_g \approx 23$ aF was observable up to a current of about $I \approx 2$ nA (Figure 8 (b)), and the curves clearly show that the activity of rare TLF-like switching events was intensified with increasing current. The capacitance ratio of the tunnel junctions, deduced from the slope of the modulation curves (see inset), was $C_1/C_2 \approx 0.84$. From the maximum modulation $U_{max} \approx 650$ µV, according to the stability diagram, we derived $C_\Sigma \approx (250 \pm 20)$ aF.

The self-capacitance contribution $C_0$ of the rather long island strip-line on the membrane substrate was estimated on the assumption that the barrier properties were the same as for the sample previously described. Hence, with $\varepsilon_r = 4.5$ and $d = 1.4$ nm, we obtain $\approx 30$ fF/µm² for the specific junction capacitance, so that each junction with an area of about 0.0014 µm² should contribute $C_T \approx 40$ aF. Consequently, $C_0 \approx 150$ aF, or, using relation (3), $\varepsilon_{eff} \approx 2$. This seems to be a realistic value for the effective dielectric constant of the island surroundings, consisting of the thin $Si_3N_4$ substrate ($\varepsilon \approx 6 - 7$) and vacuum ($\varepsilon = 1$).

Turning to the noise behavior, we found a $1/\sqrt{f}$-like dependence of the equivalent charge noise spectra in the frequency range from 1 Hz to 100 Hz and over a large range of bias currents $10\,\text{pA} \leq I \leq 1\,\text{nA}$, as shown in Figure 9 (a). At even higher currents the charge noise signal could not be detected due to the suppression of the slope of the corresponding modulation curves. In contrast to the first transistor, prepared on bulk substrate, the current dependence of the charge noise signal of this transistor was very weak: At $f = 10$ Hz, it rose from $\approx 2.5 \times 10^{-4}$ e/$\sqrt{\text{Hz}}$ at $I = 10$ pA to only $\approx 6 \times 10^{-4}$ e/$\sqrt{\text{Hz}}$ at $I = 1$ nA. As the log-log plot in Figure 9 (b) shows, this current dependence approximately obeyed a power law $Q_{Noise} \propto I^{0.18}$. In comparison with the transistor sample on bulk substrate, these superior noise properties might be attributed to the thin membrane substrate, possibly reducing the number of charge traps in the vicinity of the island. On the other hand, a similar behavior was recently reported for an Al/AlO$_x$/Al transistor, prepared on silicon substrate with AlO$_x$ buffer layer in between.[52] Here, the noise level at 10 Hz also rose from $\approx 2.5 \times 10^{-4}$ e/$\sqrt{\text{Hz}}$ ($I = 10$ pA) to $\approx 8 \times 10^{-4}$ e/$\sqrt{\text{Hz}}$ ($I = 1$ nA) with a power law dependence $Q_{Noise} \propto I^{0.25}$. Therefore, the effective influence of the substrate thickness on the noise behavior should be clarified by further experiments.

## 5. Summary

We fabricated and analyzed the tunneling properties of Cr/CrO$_x$/Cr single junctions and SET transistors, prepared by a direct-writing multilayer technique on Si/SiO$_2$ bulk and Si$_3$N$_4$ membrane substrate materials. The chromium oxide barriers were formed by oxidation in air under ambient conditions. The junction areas of the two SET transistors were 100 x 100 nm² and 17 x 60 nm², and the transistor islands were about 6.5 µm and 10 µm long.

Measurements on our single junctions with areas of about 0.5 x 0.5 µm² and 1 x 1 µm² confirmed earlier observations[21,22] that the tunneling characteristics of Cr/CrO$_x$/Cr junctions are dominated by an anomaly in the low bias regime. The physical mechanism for this behavior is not yet understood at present. In the higher voltage range, fitting of the Simmons model to the measured IVCs showed reasonable agreement, revealing that the tunneling current was dominated by the barrier suppression at $|U| \approx \phi/e$. The deduced barrier thickness was about 1.4 nm, and the barrier height $\phi \approx 740$ meV is about twice to three times smaller than the value for traditional Al/AlO$_x$/Al junctions.

Both transistor samples, prepared on conventional Si/SiO$_2$ ($C_\Sigma \approx 820$ aF) and on Si$_3$N$_4$ substrate ($C_\Sigma \approx 250$ aF), showed but good uniformity of their tunnel junctions and a moderate charge noise level ($\approx 2.5 \times 10^{-4}$ e/$\sqrt{\text{Hz}}$ at $f = 10$ Hz). A special feature of the sample on the membrane substrate was its weak current dependence of the noise. In conclusion, the noise levels obtained and their current dependencies were comparable to those of conventional Al/AlO$_x$/Al SET transistors. Further improvement of the noise behavior might be achieved by reduction of the island length.

Generally, as our results demonstrate, in most cases the inherently non-linear IVCs of SET devices with Cr/CrO$_x$/Cr junctions should be no principal disadvantage for their application. Unfortunately, one exception is the field of Coulomb blockade thermometry (CBT),[55] where the property of chromium of maintaining normal conductivity even at very low temperatures could be advantageous since the quenching of superconductivity by an external magnetic field would not be necessary. But, due to the strong, not SET-based non-linearity around zero bias, the applicability of chromium-based tunnel arrays for CBT is obviously excluded.

Despite the convincing performance of our all-chromium SET transistors, some open questions remain: First, at high bias voltages, instabilities showed up in the IVCs of the single junctions and the transistor samples, which are attributed to impurities in the barrier. Secondly, the results of our fits for the single junctions imply that the effective tunnel junction areas are smaller than expected from their geometrical sizes. Furthermore, the value for the specific tunnel capacitance, estimated from the experiments on the transistor samples, shows that the effective dielectric constant of the barrier material differs from the (expected) value of Cr$_2$O$_3$. Thus, the barrier composition and the interface quality should be examined in more detail.

We think that the presented direct-writing technique in combination with the use of the chromium material system offers considerable benefit for the preparation of special SET circuits. For instance, the multilayer technique allows the implementation of SET circuits with complicated layout features, such as devices with three-dimensionally shaped islands, traversing the gate electrodes underneath like suspended bridges, and so effectively increasing the capacitive coupling between island and gate electrode.[13] In future, this property might be important for the realization of SET devices where the cross-capacitance between different islands and "foreign" gate electrodes should be minimized, as for the realization of SET turnstiles or pumps. The main advantage of the special STEM lithography is its very high resolution that might open the door to a direct writing EBL-technique for feature sizes smaller than 10 nm, as demonstrated by recent resolution tests.[54] Also, the use of very thin membrane substrates might reveal advantages with respect to the reduction of background charge noise aimed at.

**Acknowledgments**

The authors are pleased to thank R. Dolata, S.V. Lotkhov, Yu.A. Pashkin and G. Lilienkamp for valuable discussions, F.J. Ahlers for his support with the cryogenic systems, and U. Becker for technical help with the electronic setup. This work was financially supported by the German BMBF (Grant No. 13N6260) and the EU (SMT Research Project SETamp).

**Tables**

| Sample | √A in µm | √A* in µm (± 0.05 µm) | d in nm (± 0.05 nm) | φ in meV (± 50 meV) |
|---|---|---|---|---|
| Chip 35 | 0.53 | 0.15 | 1.40 | 740 |
| Chip 44 | 1.0 | 0.42 | 1.35 | 740 |

Table 1: Results for the fitting parameters $A^*$ (effective junction area), $d$ (effective barrier thickness), and $\phi$ (effective barrier height) for Cr/CrO$_x$/Cr junctions with different geometrical areas $A$.

**Figure captions**

Figure 1: IVCs of the Cr/CrO$_x$/Cr single tunnel junction, recorded at different temperatures. The regions at high bias, where switching events in the IVCs occurred, are indicated. The inset shows a plot of the differential resistances. The typical "anomaly" of Cr/CrO$_x$/Cr tunnel junctions appears as a peak around zero bias, sharpened at low $T$.

Figure 2: Measured IVC (straight line) together with the fit of the Simmons formula (1) (dashed line), and the plot of their dynamic resistance (inset). Above $U \approx 0.4$ V, the curves are in good agreement.

Figure 3: Fowler-Nordheim plot of the IVC of a Cr/CrO$_x$/Cr single junction. The slope for the field-emission regime, corresponding to the parameters from the fit, is indicated by the dotted line. The short dashed line corresponds to the slope of the curves at maximum bias (see text).

Figure 4: (a) SEM image of a Cr/CrO$_x$/Cr SET transistor, prepared by direct-writing multilayer technique on Si/SiO$_2$ substrate. Source, drain and gate electrodes are made in the first lithographic lift-off step, followed by oxidation and deposition of the crossing island line. (b) Blowup of a tunnel junction with an area of about 100 x 100 nm$^2$.

Figure 5: (a) IVCs of the Cr/CrO$_x$/Cr SET transistor in the small bias region, recorded at different values of gate voltage $V_g$. The inset shows the typically nonlinear IVC in a large bias range. (b) Gate modulation curves of the bias voltage $U$, recorded at different bias currents, with a period corresponding to a gate capacitance of 52.5 aF. The modulation curves for very small currents ($I = \pm 5$ pA) approximate the shape of the stability diagram (gray rhomb in the inset), yielding $C_\Sigma \approx 820$ aF.

Figure 6: (a) Voltage noise spectra (rms) of the transistor for 1 Hz $< f <$ 100 Hz, measured in the current bias mode in different working points of the $U(V_g)$ modulation curve (see inset). (b) Equivalent charge noise spectra for different bias currents. The dashed lines indicate $1/f$- and $1/\sqrt{f}$-dependencies.

Figure 7: SEM image of the Cr/CrO$_x$/Cr SET transistor prepared on a Si$_3$N$_4$ membrane. The bottom electrodes (60 nm wide) were made using a "conventional" EBL system. The perpendicularly crossing island line (about 17 nm wide and 10 µm long) was exposed using the STEM. The blowup in the inset shows the area of a tunnel junction.

Figure 8: (a) IVCs of the SET transistor shown in Figure 7 for different gate voltages. (b) Gate modulation curves for different bias currents, measured at $T = 8$ mK ($C_g = 23$ aF). The inset shows modulation curves for $I = \pm 10$ pA that approximate the shape of the stability diagram (gray rhomb), yielding $C_\Sigma \approx 250$ aF.

Figure 9: (a) Equivalent charge noise spectra (rms) of the transistor shown in Figure 7, measured at different bias currents in the working points of maximum charge sensitivity. The dashed line indicates $1/\sqrt{f}$-dependence. (b) Current dependence of the equivalent charge noise signal at $f = 10$ Hz (log-log plot).

**Figures**

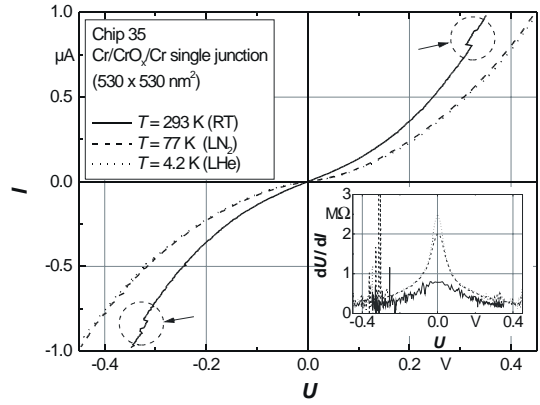

Fig. 1

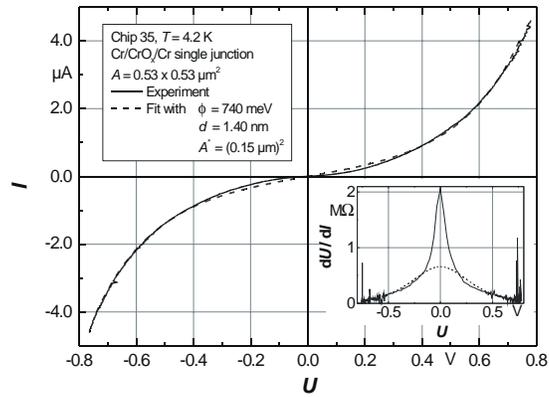

Fig. 2

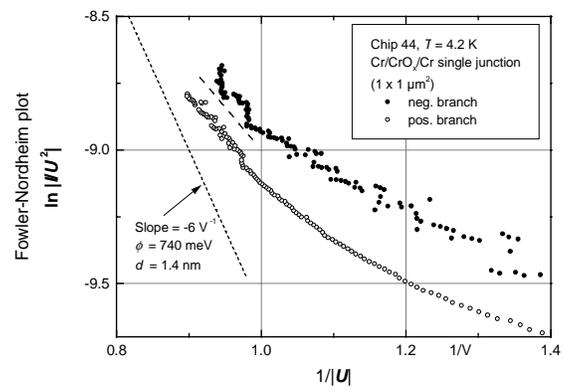

Fig. 3

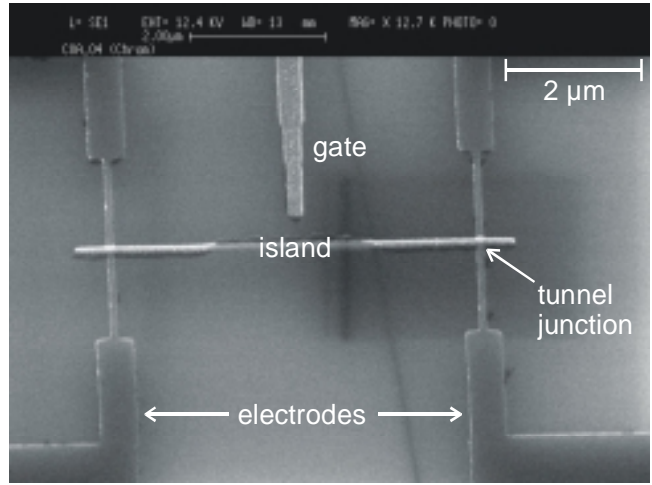

(a)

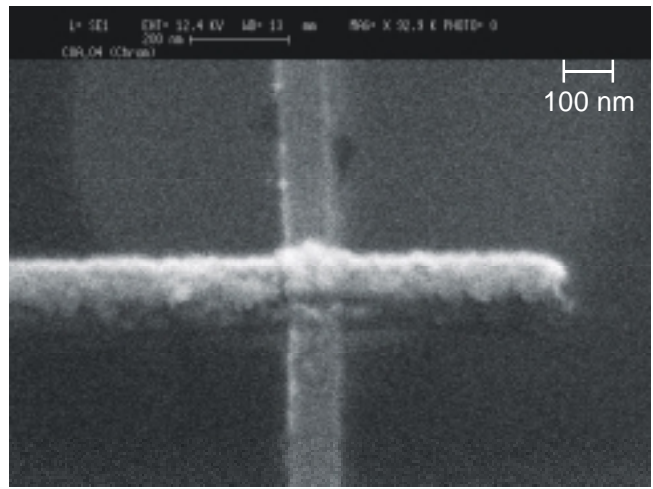

(b)

Fig. 4

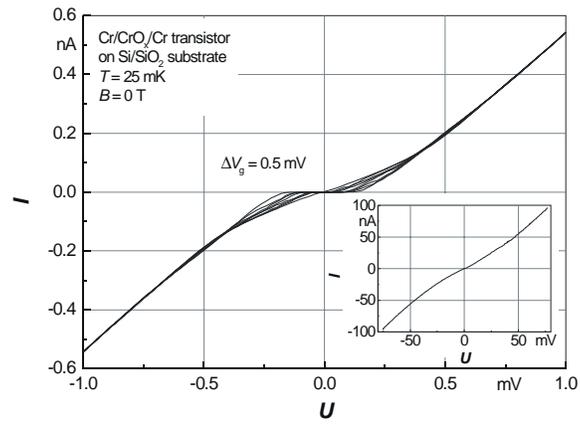

(a)

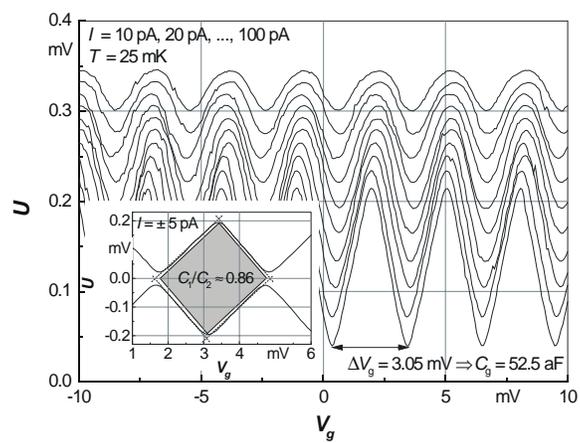

(b)

Fig. 5

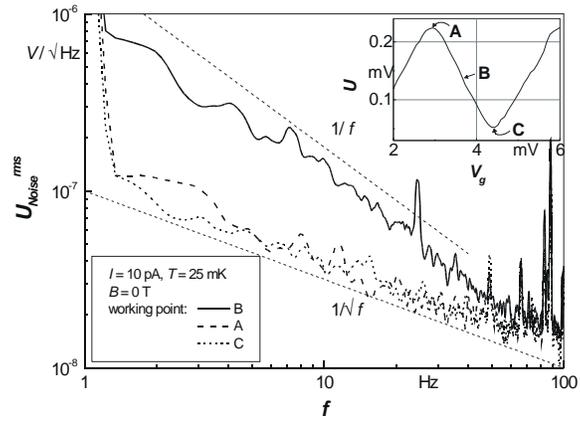

(a)

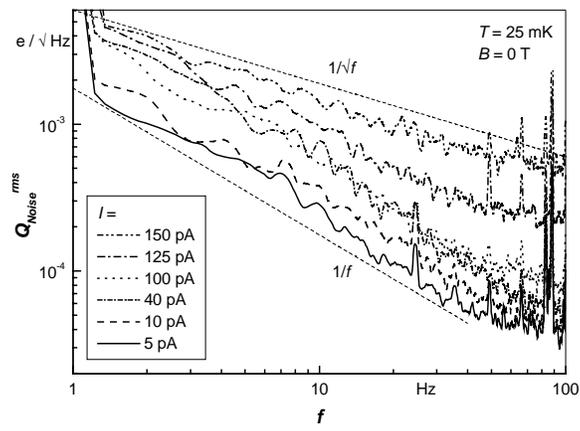

(b)

Fig. 6

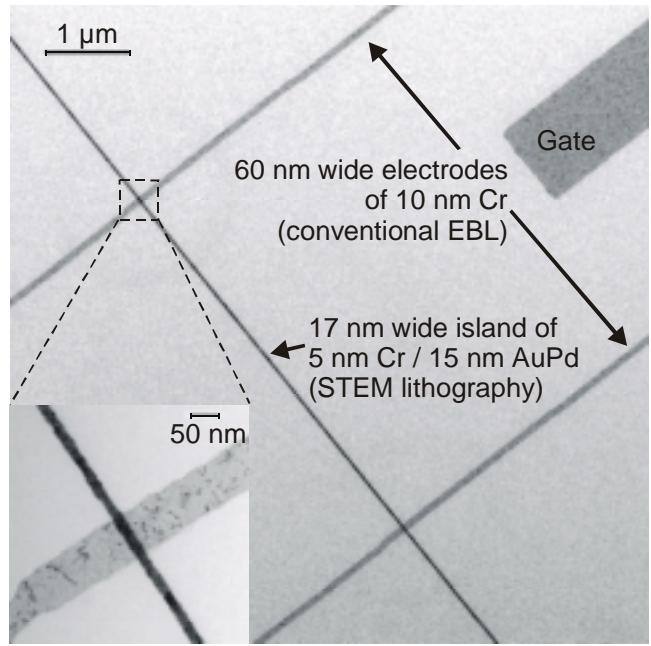

Fig. 7

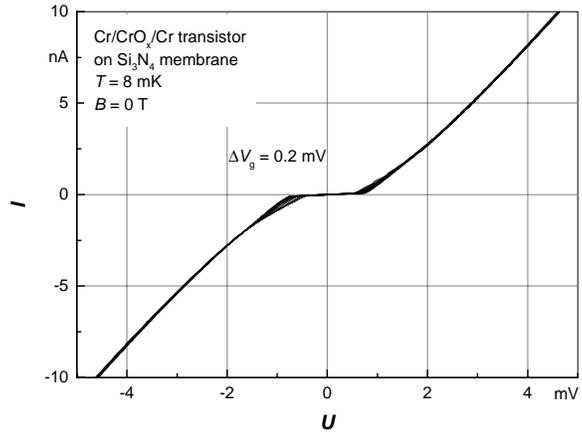

(a)

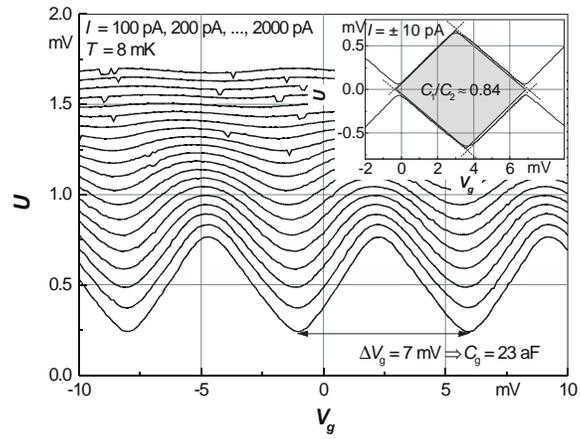

(b)

Fig. 8

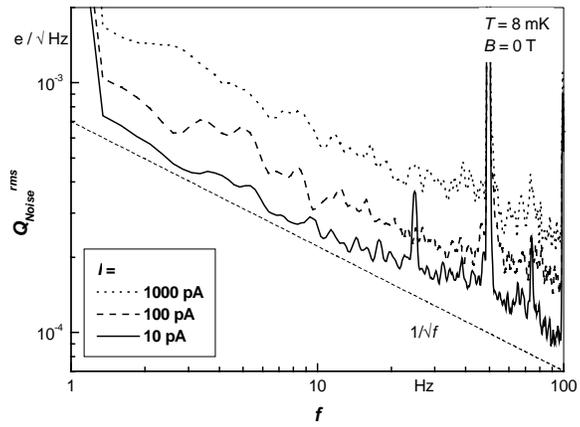

(a)

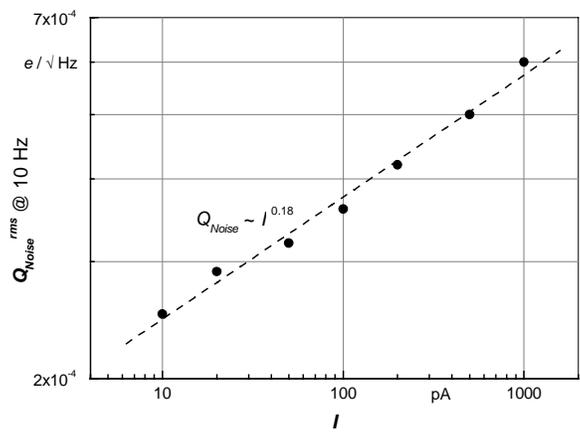

(b)

Fig. 9
19